# MyLibrary: A Model for Implementing a User-centered, Customizable Interface to a Library's Collection of Information Resources

*Eric Lease Morgan*

Digital Library Initiatives
Box 7111, NCSU Libraries, Raleigh, NC 27695-7111, USA
Tel: (919) 515-4221
E-mail: eric_morgan@ncsu.edu

**ABSTRACT**
The paper describes an extensible model for implementing a user-centered, customizable interface to a library's collection of information resources. This model, called MyLibrary, integrates the principles of librarianship (collection, organization, dissemination, and evaluation) with globally networked computing resources creating a dynamic, customer-driven front-end to any library's set of materials. The model supports a framework for libraries to provide enhanced access to local and remote sets of data, information, and knowledge. At the same, the model does not overwhelm its users with too much information because the users control exactly how much information is displayed to them at any given time. The model is active and not passive; direct human interaction, computer mediated guidance and communication technologies, as well as current awareness services all play indispensable roles in this system.

**KEYWORDS**: digital libraries, interactive assistance, librarianship, MyLibrary, MyLibrary@NCState.

**BACKGROUND**
Traditionally, libraries concentrated on the selection and storage of book and journal collections. This was true because books and journals were a primary manifestation of data, information, and knowledge. Consequently, librarians spent much of their time managing book and journal collections. Now, with the advent of globally networked computers, information increasingly appears in digital form and is often times "born digital". Library's will continue to support their print-based collections. At the same time, the libraries must be dedicated to exploring methods and facilitating access to the wealth of available digital information. MyLibrary, the model described here, is one method for addressing these issues.

In the Fall of 1997 the NCSU Libraries' Digital Library Initiatives Department conducted focus group interviews with a number of NC State students and faculty. One of the more common themes articulated by focus group participants was the desire to have access to the total universe of information but to display only the information needed for focused study and research. They felt they were "drinking from the proverbial fire hose".

At the same time services like My Excite, My Yahoo, and My DejaNews had become available. These services allow users to create profiles representing their information needs. Unlike libraries, these services do not provide access to scholarly materials; their content is very much like the content of a newspaper.

More importantly, these sites do not provide access to any sort of interactive assistance. There are no human beings available at these sites, "portals" as they are called. There are no "reference assistants" online to help with particular information queries. The only kind of assistance provided by these sites are the help texts that go unread by most users.

It was from within this environment where the MyLibrary model was created. It is an environment where people desire access to the total universe of information, but only to individual parts of it at any one time. It is an environment where people increasingly view an information world through a Web browser, but still desire the help and advice of other people when it comes to satisfying information needs. The balance of this text describes MyLibrary in terms of this service model, it's technical requirements, and a particular implementation by the NCSU Libraries.

**MYLIBRARY MODEL**
MyLibrary is an amalgamation of the essential service components of libraries and the wealth of digital data, information, and knowledge available in a globally networked computing environment. Any collection of data, information, or knowledge that does not include some sort of assistance or instructions for the people intended to use the collection cannot be called a library. The success of a library can be measured by how well people garner knowledge (not necessarily data nor information) from its collection.

### Data store

The MyLibrary model begins with the assumption that knowledge exists and is obtainable; philosophic skepticism is denied. Furthermore, it is assumed that knowledge is the result of a processing and internalization of information where information is the organization and assignment of value to data, and data is equated with simple statements or facts. The service components of libraries provide a means to facilitate this processing.

Data and information (content) must be organized and presented in a manner comprehensible to the people for whom the data and information were intended. Content must be set within a cognitive framework that is easily communicated through language or some other common medium of expression. Without a cognitive framework data and information will wallow in a morass of meaningless numbers, symbols, signs, and words. Even if the cognitive framework used to organize the data and information is easily understandable, the framework will go unused unless it can be described using terms, phrases, and presentation modes shared between the organizer of content and its intended users.

Up to this point the model described is a rudimentary data/information store. Some people and institutions have mediated access to data/information stores through networked computers and called these systems "digital libraries". They represent the present state of digital library initiatives, but they lack the essential components defining them as true libraries, namely: 1) librarians (people) who practice, articulate, and implement library policies, and 2) interactive assistance. With the additions of these essential components, the model is transformed from a simple store into a real library, and the MyLibrary model emerges.

### Library processes

Information only exists after data has been organized and given value. The process of organizing information is one of many library processes. Other processes include the collection, storage, dissemination, and evaluation of data, information, and knowledge. The people who do this organizing can be called librarians. People with academic degrees in librarianship are far from having a monopoly on these processes; generally speaking, people with academic degrees in librarianship are people who have made a career of putting these processes into practice in libraries.

In order to effectively organize data sets into information and thus begin the creation of digital libraries, librarians must articulate, and continually re-articulate library policies. These

---

**Interactive assistance**

|  | Proactive | Reactive |
|---|---|---|
| **Human** | • newsletters<br>• site visits<br>• bibliographic instruction | • reference interviews<br>• telephone reference<br>• email reference<br>• video conferencing/relay chat |
| **Computer** | • current awareness services<br>• messages of the day<br>• browsable collections<br>• prescribed resource listings<br>• email announcements<br>• expert systems | • online help<br>• customizable interfaces<br>• searchable databases |

**Combinations of interactive assistance with example services implemented in MyLibrary**

policies outline the purpose and scope of the library, as well as determine for whom the library serves. It is then the responsibility of librarians to choose from available technologies the methods for implementing these policies. The current trend in librarianship is to use computer-based technologies for this purpose. MyLibrary is no exception. MyLibrary has been defined as a tool designed to help people access the wealth of digital information available in a globally networked computer environment and at the same time not overwhelm these people with unneeded information. MyLibrary uses an amalgamation of computer technologies (described later) to accomplish this goal.

**Interactive assistance**

The final ingredient in the MyLibrary model is "interactive assistance", a function of a library providing methods for interpreting and customizing its content. No matter how well an information system (library) is designed, there are always to going to be people who cannot locate the information they seek even if it exists within the system. The purpose of interactive assistance is to reduce this possibility by providing specialized help for specialized situations and the means for restructuring a library's collection to fit each individual's needs. Interactive assistance adds a level of intelligence to the use of MyLibrary.

Interactive assistance can be proactive or reactive. Proactive interactive assistance queries users for their information needs. It analyses the answers to the queries and either formulates possible solutions or continues the query process. A reactive assistance approach only provides possible solutions after being asked questions by users. The difference between these two approaches is similar to the difference between browsability and searchability. Both browsability and proactive assistance layout ready-made solutions or information paths. Searchability and reactive assistance require the users to articulate information needs and translate them into the language of the system. Like browsability and searchability, ideally, elements of both proactive and reactive assistance are desirable in the implementation of any library or information system.

Interactive assistance can take place through direct human communication channels or it can be mediated through technology (computers). The traditional "reference interview" is seen as the most effective example of direct human communication for interactive assistance. It is supplemented with non-verbal signals (like facial expressions), the process is customized in each session, and more importantly, the feedback is immediate. At the same time, direct human communication is time intensive, and since the people trained in reference interview techniques are few in number, direct human communication is not scalable. Computer mediated assistance is asynchronous in nature where the people (librarians) maintaining the information system (library) are at least one step removed from users using the system. Computer mediated interactive assistance services have the ability to reach a wider audience of users, but

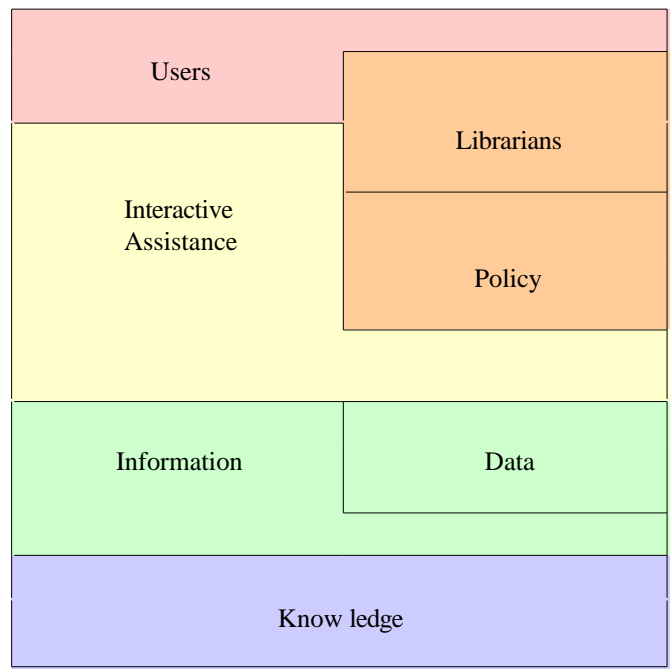

**The service model of MyLibrary**

generally speaking, the services are not personalized for each individual. Artificial intelligence and expert systems aspire to fulfill computer mediated interactive assistance, but they have yet to provide long-term solutions to users' demands.

Using the model described above the NCSU Libraries has begun implementing a MyLibrary system. It incorporates every element of the model described so far. It includes at least one combination of the proactive/reactive and human/computer interactive assistance services. The implementation is called MyLibrary@NCState. The balance of this paper describes the interfaces to MyLibrary@NCState emphasizing how they exemplify elements of the MyLibrary service model. It also describes the technical infrastructure supporting these interfaces. The paper concludes with an outline of future enhancements for the system.

### INTERFACES TO MYLIBRARY@NCSTATE

MyLibrary@NCState is an implementation of the MyLibrary service model at the NCSU Libraries of the North Carolina State University in Raleigh, NC, USA. This section describes the user and administrator interfaces to the implementation.

#### User Interface

From the user's perspective, MyLibrary@NCState is a customizable HTML page. It is accessible via any Web browser supporting Netscape cookies and the Secure Socket Layer (SSL) protocol. The process of creating a MyLibrary@NCState page begins when the user creates an account on the system. The account information includes rudimentary biographical data: name and email address. In addition, the creation of an account necessitates the selection of a primary academic interest area from a pop-up menu. Once an account is created, the user's MyLibrary@NCState page is dynamically built and displayed. Afterwards other links may be added to or subtracted from the preconfigured list of recommendations creating the framework for the user's personal "digital library".

For example, by selecting among the system's many Customize hotlinks, a user is presented with lists of disciplines. Each list is associated with information resources (databases, journals, Internet resources, etc.) specific to that discipline. Resources from any discipline may be chosen to be included in the user's MyLibrary@NCState personalized digital library, which appear in the Web browser after the selections are submitted. The next time a user visits MyLibrary@NCState, the system remembers all customizations and lists them accordingly.

Resources on the user's MyLibrary@NCState page include information about the system, messages from librarians, links to a user's personal librarians, university resources, discipline-specific Internet resources, citation databases, electronic journals, as well as direct access to common search engines and an NCSU Libraries-based selective dissemination of information service. These resources are divided into the sections below. Items marked with a dagger are customizable:

- header
- global message
- Message from the Librarian
- Your Librarians
- Library Links †
- University Links †
- Current Awareness †
- Personal Links †
- Quick Searches †
- Reference Shelf †
- Bibliographic Databases †
- Electronic Journals †
- footer

The purpose and functions of each of these sections are described below.

*Header and Footer.* The header displays the MyLibrary@NCState logo, a navigation bar, a customize link, and a logout function. The logo gives the service an identity. The navigation bar provides easy movement around the page. The customize link initiates the process for changing the page's content. The logout function removes the system's Netscape cookie from the user's computer allowing the user to access the service at public workstations and at the same limiting unauthorized access to a user's page. The footer displays version, date, and contact information for the system. The user cannot customize the header nor the footer.

*Global message and Message from the Librarian.* These services are proactive/computer mediated interactive assistance functions. They provide the means for librarians to broadcast messages to the users of MyLibrary. The global message is a text intended for every user of the system. It is much like the message of the day (MOTD) function on Unix computers. Information intended for the global message function includes announcements in changes of service or downtime, news from around the campus, or simply an interesting URL from the University's newspaper.

The Message from the Librarian is intended to function exactly like the global message except its content is only displayed to people who have chosen particular disciplines. Consequently there are many Messages from the Librarians, but not more than the total number of disciplines. Every user of MyLibrary@NCState is associated with an academic discipline. Each of these disciplines is associated with a text

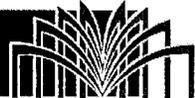

**A screen dump of one user's personalized MyLibrary@NCState page.**

message intended to be changed regularly. When a MyLibrary@NCState page is displayed a discipline-specific message for the user is included in the content. These messages are expected to contain information pertinent to the discipline like URLs of interest, announcements of training opportunities or the acquisition of new resources, or a short description of library-related services of interest. The user cannot customize either of these message functions.

*Your Librarians.* The Your Librarians section lists the names, telephone numbers, and email address of the librarians and collection managers associated with the user's chosen discipline. Because each user of MyLibrary@NCState is associated with a discipline, and since each discipline is associated with at least one reference librarian and collection manager, then each user can be made aware of the names and contact information of their discipline-specific librarians and collection managers. This section also lists a generic email address intended for reference questions. The Your Librarians section is not directly customizable by the user, but it does change if the user changes their selected discipline. This section is a reactive/human mediated interactive assistance function.

*Library and University Links.* The Library Links and University Links sections work similarly. Each is populated initially with a librarian-defined list of hyperlinks to various library and University home pages. These home pages are of any type but usually include pages describing library/university services, hours, policies, finding aids, University sites, or campus-wide directories.

These sections are customizable by the user. By selecting an associated Customize hotlink, the user is presented with a list of available library/university-related information sources deemed by the librarian as "important" or "useful to know". Each item in the list is associated with an HTML checkbox. The user selects or de-selects any number of these items, and by submitting their selections, the system records the user's preferences in the underlying database and redisplays their MyLibrary@NCState page with only the selected items.

Since these sections are initially populated by librarians who are making recommendations based on disciplines, these sections are initially proactive/computer mediated interactive assistance services. After these sections have been customized by the user, they become reactive/computer mediated interactive assistance services.

*Bibliographic Databases, Electronic Journals, and Reference Shelf.* The Bibliographic Databases, Electronic Journals, and Reference Shelf sections provide the means for the user to have displayed only the scholarly information resources they believe are important to their work.

Like the Library Links, these sections are initially populated with a list of recommendations prescribed by a discipline-specific librarian. The Bibliographic Databases section contains lists of journal indexes. The Electronic Journals section contains lists of digital serial titles. The Reference Shelf was originally intended to include links to Internet resources traditionally associated with traditional library reference desks (ie. dictionaries, encyclopedias, handbooks, manuals, directories, maps, etc.) but it has since grown to include just about any type of Internet resource with scholarly content.

To customize these sections the user selects the associated Customize hotlink. They are then presented with a list of their existing section items as well as a list of all the system's disciplines. From here the user can check or uncheck any of their existing section items and return to their MyLibrary@NCState page. Alternatively, the user can select a discipline. This returns a list of all the resources associated with that discipline and that section. The user can select any of these resources. Upon submitting their choices their selections are saved to the system's database and the user is returned to the main page.

Like the Library and University Links, these sections are initially proactive/computer mediated interactive assistance services. After these sections have been customized by the user, they become reactive/computer mediated interactive assistance services.

*Personal Links.* The Personal Links section allows the user to save links to Internet resources they use often but not included in any of the sections above. No librarian can hope to collect the total sum of Internet resources and organize them into a database. Given this fact, the Personal Links section provide the opportunity for the MyLibrary@NCState user to add their own links to the system.

To customize this section the user first selects the Customize hotlink. They are then presented with a list of the existing personal links and given the opportunity to delete and of them or add others. Upon submitting the customization form the user is returned to the main MyLibrary@NCState page. This section is represents a reactive/computer mediated interactive assistance function.

*Quick Searches.* The Quick Searches section provides direct access to a number of Internet search engines including local online public access catalogs (OPAC). By selecting a search engine from a pop-up menu, entering a query in the text field, and submitting the form, a database query is constructed in the form of a URL. The user's browser is then "redirected" to this dynamically created URL and the results of the search are displayed.

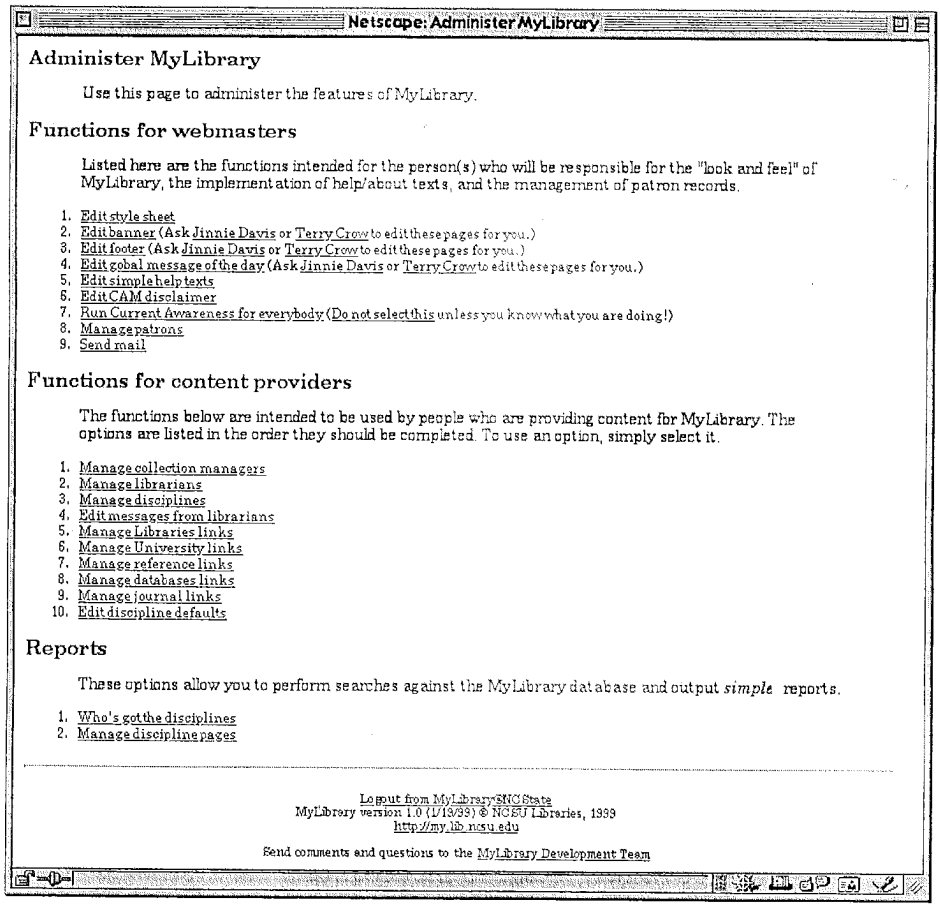

**A screen dump of the MyLibrary@NCState administrative interface.**

Customizing this section works just like the customizing process of the Library and University Links sections. It is also initially a proactive/computer mediated interactive assistance service. After it has been customized by the user, it becomes reactive/computer mediated interactive assistance service.

*Current Awareness.* The Current Awareness section provides the means for the user to search the library's OPAC for recent acquisitions in the user's interest areas. Optionally users can have these search results regularly sent to their email address. Assisted with a simplified version of the library's call number system, the user can customize this function so it saves ranges of call numbers, profiles, to the MyLibrary@NCState database. Users can save as many of these profiles as they desire. The profiles are searched on a regular basis (weekly) against a list of new acquisitions to the library's catalog, and the results sent to the user's email address. The results include only call number, author, and title information. Each item is associated with a URL allowing the user to view the full record describing that item in detail. Based on this full record, the user can then decide whether or not they want to borrow the item from the library's collection. This section is a reactive/computer mediated interactive assistance function.

### Administrative interface

Maintenance of MyLibrary@NCState system is done through an administration interface. The interface is a menu of hypertext links protected by a simple username/password access control system. The administration menu allows administrators to create, modify, or delete just about any content item in the system's database. The most important item in the menu is the list of academic disciplines since each and every other item in the system is somehow related to this list. Other important content times include:

- the names and contact information of librarians
- links and descriptions of information resources (databases, journals, etc.)
- help texts,
- global message and messages from the librarians
- the system's header and footer

To create, modify, or delete any of these items the authorized librarian selects an item from the menu and completes the resulting form. Changes take place immediately.

For example, suppose somebody wanted to change the global message. They would select the Edit Global Message of the Day hotlink and enter what ever text they desired. If a new librarian where to join the library's staff, then they would select the Manage Librarians hyperlink and create a new record making sure the record was associated with at least one discipline. Consequently, users of the system who had selected a discipline equal to the discipline chosen by the new librarian would then see the new librarian's name and email address on their MyLibrary@NCState page.

The administrative interface hosts two other functions: reports and email. The reports function is immature, but in conjunction with a standard HTTP common log file, will provide access to qualitative and quantitative data describing how the system is being used.

Using the email function, a librarian can mass email all the people in one or more disciplines. The content of these email messages is intended to be of the same sort of information as the Messages From The Librarian. Consequently, this function is a proactive/computer mediated interactive assistance service. Users do not have to receive this email unless they desire it since it is an option in each user's biographical profile allowing them to turn it off or on.

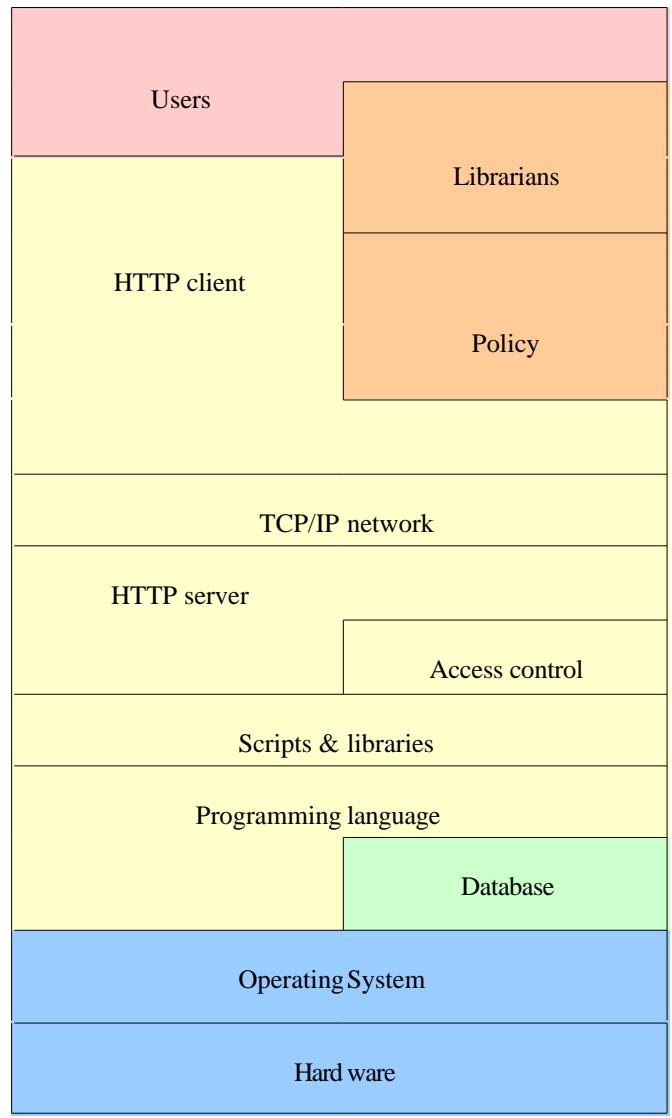

**The computing model of MyLibrary**

# MYLIBRARY@NCSTATE TECHNICAL INFRASTRUCTURE

MyLibrary@NCState is essentially a database application, accessible from an HTTP/HTML interface, and governed by the principles and practices of librarianship. The distinctive element of this system is not the technology driving it, but the interactive assistance services it provides via librarians. Interactive assistance and librarianship were outlined previously. The text below describes the system's technical infrastructure, the computing model.

MyLibrary@NCState is primarily built on four software technologies:

1. an operating system
2. a structured query language (SQL) database server
3. a hypertext transfer protocol (HTTP) server
4. a scripting language "gluing" together items #1, #2 and #3

## Operating System

MyLibrary@NCState is grounded on a computer running Unix, and just about any computer running Unix can run MyLibrary@NCState. Unix was chosen for two reasons. First, Unix is the most commonly used operating system for large Internet services. Consequently, there exists a large number of people who can administrate Unix computers. Second, Unix is available for the widest range of hardware platforms, and since the NCSU Libraries considers the possibility of sharing the MyLibrary@NCState source code, the Libraries does not want to limit what the system can run on.

## Database Server

The SQL database hosting MyLibrary@NCState includes multiple tables holding data for the majority of the system's content (user and current awareness profiles, Internet resources, help texts, recent acquisitions information, and relationships between these tables). Creation of these tables is done through a minimalistic but wholly functional terminal interface, but data entry functions are supported via HTML forms from the Administrative interface.

The database application serving all this content is MySQL. It was chosen primarily because it was free, implements SQL, runs on multiple Unix computers, provides the mechanisms for auto-incrementable fields as well as variable field lengths, but most importantly, it supports an application programmer's interface (API) for both the Perl and C programming languages.

## HTTP server

The HTTP server hosting MyLibrary@NCState is a Netscape Enterprise Server, but any HTTP server would work as long is could run on a Unix computer and support CGI scripts; MyLibrary@NCState does not rely on any special functions of the Netscape Enterprise Server.

## Scripting language

Perl is the CGI scripting language chosen to "glue" together the SQL database and the HTTP server. Perl was the obvious choice for this purpose since it is interpreted (making it easy to debug), free, works on multiple computing platforms, widely supported by the Internet community, but most importantly is one of the APIs supported by MySQL.

MyLibrary@NCState is really two sets of Perl scripts. One is for the user interface and the other is for the administrative interface. Both sets "require" and "use" many supporting subroutines. Including both POD (plain old documentation) and comments, the entire system comprises more than 13,000 lines of code.

*Program execution.* Program execution begins when a user connects to the default file (a script) of the HTTP server and initializes a number of global variables. If the user's browser sent a MyLibrary@NCState Netscape cookie, then the value of that cookie is uses as a database key to display a specific MyLibrary@NCState page. If a MyLibrary@NCState cookie was not sent, then the user's browser is redirected to the University's SSL authentication system. The user's authentication is then used as a key to locate a MyLibrary@NCState database record. If a record is not found then a new record is created and a MyLibrary@NCState cookie is returned to the user. If a record is found then a MyLibrary@NCState cookie is returned. In either case, the user then has the opportunity to view their MyLibrary@NCState page and customize it accordingly.

Customization commands are sent via the command line argument of a URL. Program execution branches to specific subroutines based on the values of these command line arguments. In general, a command line argument will be of three types: get, set, or display. Get commands, not to be confused with GET HTTP action statements, display a user's current settings and allow them to choose other options via HTML forms. Set commands take the input of get commands and save them to the system's database. If the command is not get nor set, then a user's MyLibrary@NCState page is displayed.

## NOT JUST A PORTAL OR BOOKMARK MANAGER

There are at least four things that make MyLibrary@NCState not just another portal or bookmark manager. First, the system includes a service allowing users to regularly receive and search lists of new books added to the library's collection, the Current Awareness service. Using Library of Congress call numbers for books and serials, users can create any number of Current Awareness profiles. The MyLibrary@NCState system saves these profiles and searches its database for them on a regular basis. Search results are sent to a user's email address, which allows direct access to the library's catalog via a hotlink and to more information describing the located item.

Second, based on the selected discipline, the system displays the name and contact information for the appropriate librarian and collection manager who specializes in the given subject area. More than one librarian may be listed, depending on the discipline selections. The library has learned from its users that although digital libraries are desirable, direct access to librarians is necessary as well.

Students and faculty or administrative staff do not always have time to be constantly on the lookout for new and better information resources. The library helps in this regard with another service of MyLibrary@NCState called Message from the Librarian. This section, updated regularly by the appropriate librarians announces, suggests, and helps users keep abreast of interesting information developments in their selected disciplines.

Finally, unlike bookmark managers, MyLibrary@NCState is portable and requires only a Web browser to use. Because access to the system's database is keyed to the University's authentication system, valid users can access MyLibrary@NCState from just about any Web browser in the world. MyLibrary@NCState can be accessed from offices, homes, or even within the library. Conversely, bookmark managers or locally defined sets of bookmarks are bound by a particular machine or computing system.

## MYLIBRARY AND MYLIBRARY@NCSTATE FUTURE DIRECTIONS

The MyLibrary model is sound, and the implementation of MyLibrary, MyLibrary@NCState, is stable, but this does not mean there is not room for improvement. This section outlines some future directions for the ideas described in this paper.

First of all, the MyLibrary model's definition of the relationship between data, information, and knowledge is easily debatable by philosophers, psychologists, and information scientists. At the same time, these epistemological issues have gone on unresolved since at least the birth of Western philosophy.

Second, if the epistemological plausibility of the MyLibrary model is assumed, then the question arises, "When does a user have enough data, information, or knowledge to satisfy their needs?" Put another way, how is the value of data, information, or knowledge measured? Unlike money, which can be defined as a common medium of exchange, the value of data, information, or knowledge is not consistent between people. Consequently the value of MyLibrary's content and services is very difficult to describe in terms of quantity or quality. If a means for measuring the value of MyLibrary's content and services were articulated, then those means could be put into place and ultimately support a better model.

Third, MyLibrary@NCState does not provided enough flexibility allowing users to customize it for their own needs. The current implementation cubbyholes users' information needs into a limited set of five or six groups of information (personal links, bibliographic databases, electronic journals, etc.). The implementation should allow the user to create their own groups of information. Future implementations will provided this functionality.

Fourth, MyLibrary@NCState does not support enough direct human communication channels. Users desire or require the assistance of other people when it comes to information work. If MyLibrary@NCState supported additional, synchronous modes of communication between librarians and users (or users and other users), then users might spend more time evaluating data, information, or knowledge rather than trying to locate them. Supplemental communication channels include simple relay chat rooms, telephony, or video conferencing.

It is an interesting time for librarians and other information professionals. As the global economy continues to move away from manufacturing enterprises and towards service industries, information professionals are finding an abundance of opportunities. At the same time they are realizing there are hosts of problems associated with these new opportunities. The MyLibrary model is one possible solution to some of these problems.